\documentclass[11pt,a4paper]{article}

\usepackage[cp1251]{inputenc}
\usepackage[T2A]{fontenc}
\usepackage[russian]{babel}

\usepackage[dvips]{graphicx}
\usepackage{amsmath}
\usepackage{amssymb}
\usepackage{amsxtra}
\usepackage{amsthm}
\usepackage{latexsym}
\usepackage{array}
\usepackage{multirow}
\usepackage{hhline}

\newtheorem{theorem}{Теорема}

\numberwithin{equation}{section}

\newcounter{myta}
\newcommand{\myt}{\refstepcounter{myta}\themyta}

\newcommand {\A} {\mathop{\rm Acc}\nolimits}

\newcommand {\lsgn} {\mathop{\rm bsgn}\nolimits}
\newcommand {\ri} {\mathrm{i}}
\newcommand {\mbf}[1] {\mathbf{#1}}
\newcommand {\fts}[1] {{\small #1}}
\newcommand {\ts}[1] {\textsl{#1}}
\newcommand {\bT}{\mathbf{T}}

\newcommand {\mP}{\mathcal{P}}
\newcommand {\mB}{\mathcal{B}}
\newcommand {\mD}{\mathcal{D}}
\newcommand {\bs}{\boldsymbol}

\def\mPs{\mP}

\def\gs{\geqslant}
\def\ls{\leqslant}
\def\bR{\mathbb{R}}

\newcommand {\ds}{\displaystyle}
\begin{document}

\begin{center}

\Large{{\bf Разделение переменных и бифуркации
первых интегралов в одной задаче Д.Н.
Горячева}}

\vspace{5mm}

\normalsize

{\bf П.Е.\,Рябов}\footnote{Работа выполнена при
финансовой поддержке гранта РФФИ и АВО №
10-01-97001.}

\vspace{4mm}

\small

Финансовый университет при Правительстве РФ

Россия, Москва

E-mail: orelryabov@mail.ru
\end{center}

\begin{flushright}
{\it Получено ZZZZZZZZZZZZZZZZZZZZZZZZZZ}
\end{flushright}

\vspace{3mm}

{\footnotesize Построено вещественное
разделение переменных для одного частного
случая интегрируемости Д.Н. Горячева в задаче о
движении твердого тела в жидкости. Получены
уравнения Абеля--Якоби с многочленом шестой
степени под радикалом. Все фазовые переменные
алгебраически выражены через разделенные
переменные. На основе явных решений исследована фазовая топология задачи.

}

\normalsize

\vspace{3mm} Ключевые слова: уравнения
Кирхгофа, интегрируемая система, алгебраическое
разделение переменных

\vspace{6mm}

\begin{center}

\large

{\bf P.E.~Ryabov}

{\bf The  separation of variables  and
bifurcations of first integrals in one problem
of D.N.~Goryachev}

\normalsize
\end{center}

\vspace{3mm}

{\footnotesize
The equations of
motion in one partial integrable case of
D.N.\,Goryachev in the rigid body dynamics are
separated by the real change of
variables. We obtain the Abel--Jacobi equations with the polynomial of degree 6 under the radical. The natural
phase variables (components of the momentum and
the momentum force) are expressed via separated variables explicitly in elementary algebraic
functions. Basing on these expressions the phase topology of the system is described.}

\vspace{3mm} Keywords: Kirchhoff equations,
integrable system, algebraic separation of
variables

Mathematical Subject Classification 2000:
70E17, 70G40

\tableofcontents
%\newpage
%\small

\section{Введение}\label{sec1}
Уравнения Кирхгофа движения твердого тела в
жидкости в общем случае имеют вид
\begin{equation}\label{xx1_1}
\dot {\boldsymbol M}={\boldsymbol
M}\times\frac{\partial H}{\partial {\boldsymbol
M}}+{\boldsymbol\alpha}\times\frac{\partial
H}{\partial {\boldsymbol\alpha}},\quad
\dot{\boldsymbol\alpha}={\boldsymbol\alpha}\times\frac{\partial
H}{\partial {\boldsymbol M}},
\end{equation}
где ${\boldsymbol M}\in \bR^3$ -- импульсивный
момент, ${\boldsymbol \alpha}\in \bR^3$ --
импульсивная сила, $H=H({\boldsymbol
M},{\boldsymbol\alpha})$ -- полная энергия.
Известными интегралами системы (\ref{xx1_1})
являются геометрический интеграл
\begin{equation}\label{xx1_2}
\Gamma=\alpha_1^2+\alpha_2^2+\alpha_3^2,
\end{equation}
интеграл площадей
\begin{equation*}\label{xx1_3}
L=M_1\alpha_1+M_2\alpha_2+M_3\alpha_3
\end{equation*}
и полная энергия $H$. На совместном уровне
\begin{equation*}\label{xx1_4}
\mP_{a,\ell}=\{\Gamma=a^2,\quad L=\ell\},\quad
\dim\mP_{a,\ell}=4
\end{equation*}
система (\ref{xx1_1}) гамильтонова с двумя
степенями свободы, в связи с чем для ее
интегрируемости достаточно в дополнение к
интегралу $H$ указать еще один интеграл,
независимый с $H$ почти всюду.

В работе \cite{gory} найден случай
интегрируемости, в котором
\begin{equation}\label{xx1_5}
H=\frac{1}{2}(M_1^2+M_2^2)+M_3^2+\frac{1}{2}[c(\alpha_1^2-\alpha_2^2)+
\frac{b}{\alpha_3^2}].
\end{equation}
В предположении
\begin{equation}\label{xx1_6}
L=0
\end{equation}
система (\ref{xx1_1}) на $\mP_{a,0}$ имеет
первый интеграл
\begin{equation}\label{xx1_7}
F=[M_1^2-M_2^2-\frac{b(\alpha_1^2-\alpha_2^2)}{a^2\alpha_3^2}+c\alpha_3^2]^2+
4[M_1M_2-\frac{b\alpha_1\alpha_2}{a^2\alpha_3^2}]^2.
\end{equation}
В частном случае $b=0$ этот интеграл найден
С.А. Чаплыгиным \cite{chap}. Там же выполнено и
сведение задачи при $b=0$ к эллиптическим
квадратурам. Дальнейшие обобщения
рассматриваемая интегрируемая система получила
в работе Х.М. Яхья \cite{yehi}. Явного сведения
случая Горячева (\ref{xx1_5})-(\ref{xx1_7})  к
квадратурам до настоящего момента найдено не
было. В работе \cite{tsig} на основе идей
бигамильтонова подхода предложены переменные
разделения, однако предложенные уравнения типа
Абеля-Якоби полностью в явной форме не
выписаны. Не найдены и зависимости фазовых
переменных от переменных разделения, в связи с
чем результаты \cite{tsig} еще далеки от
завершения.

В настоящей публикации представлено явное
разделение переменных для случая Горячева. Это
решение не требует привлечения каких-либо
математических теорий. Получены простые
разделенные дифференциальные уравнения, все
исходные фазовые переменные алгебраически
выражены через переменные разделения.

\section{Параметризация интегральных многообразий}\label{sec2}
В этом параграфе мы обобщаем замены С.А.
Чаплыгина \cite{chap}, следуя геометрическому
подходу к разделению переменных, предложенному
в \cite{khsh, KhND06, khmp}.

Вместо одного из интегралов $H,F$ можно
рассматривать интеграл в форме, предложенной в~\cite{tsig}:
\begin{equation*}\label{xx2_1}
K=[M_1^2+M_2^2+\frac{b}{\alpha_3^2}]^2+2c\alpha_3^2(M_1^2-M_2^2)+c^2\alpha_3^4.
\end{equation*}
При условии (\ref{xx1_6}) он выражается через
$H$ и $F$ следующим образом:
\begin{equation}\label{xx2_2}
K=F+\frac{4b}{a^2}H-\frac{b^2}{a^4}.
\end{equation}

Подходящим выбором единиц измерений и
направления подвижных осей можно добиться того,
чтобы было выполнено $c=1$ и $a=1$.

Введем переменные $x$, $z$, $\xi$, полагая
\begin{equation}\label{xx2_3}
x^2=\alpha_1^2+\alpha_2^2,\quad
z=\alpha_3^2,\quad
\xi=M_1^2+M_2^2+\frac{b}{\alpha_3^2}.
\end{equation}
Эти переменные подсказаны работой \cite{chap},
в которой при $b=0$ они привели к разделению.
Следуя С.В.\,Ковалевской, введем также
комплексную замену ($\ri^2=-1$)
\begin{equation}\label{xx2_4}
\begin{array}{ll}
w_1=M_1+\ri M_2,& w_2=M_1-\ri M_2,\\
x_1=\alpha_1+\ri \alpha_2,& x_2=\alpha_1-\ri
\alpha_2.
\end{array}
\end{equation}
Геометрический интеграл примет вид
\begin{equation}\label{xx2_5}
x^2+z=1,
\end{equation}
а поскольку сама постановка задачи предполагает
$\alpha_3\ne 0$, из (\ref{xx1_2}),
(\ref{xx1_6}), (\ref{xx2_4}) выразим
\begin{equation}\label{xx2_6}
M_3=-\frac{x_2w_1+x_1w_2}{2\sqrt{z}}.
\end{equation}
Подстановка (\ref{xx2_3}), (\ref{xx2_6}) в
уравнения
\begin{equation}\label{xx2_7}
H=h,\quad K=k
\end{equation}
интегрального многообразия
$J_{h,k}\subset\mP=\mP_{1,0}$ приводит их к
виду
\begin{equation}\label{xx2_8}
w_1^2+w_2^2=\frac{k-\xi^2}{z}-z,
\end{equation}
\begin{equation}\label{xx2_9}
x_1^2(w_2^2+z)+x_2^2(w_1^2+z)=4hz-2\xi-2b\left(1-\frac{1}{z}\right).
\end{equation}

Из определения (\ref{xx2_3}) с использованием
(\ref{xx2_5}), (\ref{xx2_8}) имеем также
\begin{eqnarray}
& \ds{w_1w_2=\xi-\frac{b}{z}},\label{xx2_10}\\[3mm]
& z^2(w_1^2+z)(w_2^2+z)=kz^2-2b\xi
z+b^2,\label{xx2_11}
\\[3mm]
& x_1 x_2=1-z.\label{xx2_12}
\end{eqnarray}
Из (\ref{xx2_8}), (\ref{xx2_10}) находим
\begin{equation*}\label{xx2_13}
w_1=\frac{1}{2\sqrt{z}}(\sqrt{\varphi_{-}}+\sqrt{\varphi_{+}}),\quad
w_2=\frac{1}{2\sqrt{z}}(\sqrt{\varphi_{-}}-\sqrt{\varphi_{+}}),
\end{equation*}
где
\begin{equation*}\label{xx2_14}
\varphi_{\pm}(z,\xi)=k\pm 2b-(\xi\pm z)^2.
\end{equation*}
Введем комплексно сопряженные переменные
\begin{equation}\label{xx2_15}
\mu_1=z(w_1^2+z),\quad \mu_2=z(w_2^2+z),
\end{equation}
и пусть $\mu^2 = \mu_1 \mu_2$. Тогда уравнение
(\ref{xx2_11}) примет вид
\begin{equation}\label{xx2_16}
\mu^2=kz^2-2b\xi z+b^2.
\end{equation}
Отсюда, в частности, следует, что любая
траектория системы изображается в виде кривой
на поверхности второго порядка (\ref{xx2_16}) в
вещественном трехмерном пространстве $ \bR
^3(z,\xi,\mu)$.

Из (\ref{xx2_15}), (\ref{xx2_8}) находим
\begin{equation*}\label{xx2_17}
\mu_1+\mu_2=z^2-\xi^2+k.
\end{equation*}
Обозначая $\psi=z^2-\xi^2+k$,
$\psi_{\pm}=\psi\pm 2\mu$, можем записать
\begin{equation}\label{xx2_18}
\begin{array}{l}
\sqrt{\mu_1}=\ds{\frac{1}{2}\left(\sqrt{\psi_{+}}+\sqrt{\psi_{-}}\right)},\qquad
\sqrt{\mu_2}=\ds{\frac{1}{2}\left(\sqrt{\psi_{+}}-\sqrt{\psi_{-}}\right)}.
\end{array}
\end{equation}
Обозначим
\begin{equation*}\label{xx2_19}
\theta=2hz^2-(\xi+b)z+b,\quad
\theta_{\pm}=\theta\pm(1-z)\mu.
\end{equation*}
Тогда из (\ref{xx2_9}), (\ref{xx2_12}),
(\ref{xx2_15}) получим
\begin{equation*}\label{xx2_20}
\begin{array}{l}
x_1=\ds{\frac{1}{\sqrt{2
\mu_2}}\left(\sqrt{\theta_{+}}+\sqrt{\theta_{-}}\right)},\qquad
x_2=\ds{\frac{1}{\sqrt{2
\mu_1}}\left(\sqrt{\theta_{+}}-\sqrt{\theta_{-}}\right)},
\end{array}
\end{equation*}
или, с учетом (\ref{xx2_18})
\begin{equation*}\label{xx2_21}
\begin{array}{l}
x_1=\ds{\sqrt{2}\frac{\sqrt{\theta_{+}}+\sqrt{\theta_{-}}}{\sqrt{\psi_{+}}-
\sqrt{\psi_{-}}}},\qquad
x_2=\ds{\sqrt{2}\frac{\sqrt{\theta_{+}}-\sqrt{\theta_{-}}}{\sqrt{\psi_{+}}+\sqrt{\psi_{-}}}}.
\end{array}
\end{equation*}
Выражения для $M_3, \alpha_3$ находим теперь из
(\ref{xx2_6}) и второго соотношения
(\ref{xx2_3}). Таким образом, найдены все
алгебраические выражения для фазовых
переменных через две вспомогательных переменных $z,\xi$. Возвращаясь к вещественным
компонентам фазового вектора, имеем
\begin{equation}\label{xx2_22}
\begin{array}{l}
M_1=\ds{\frac{1}{2}\sqrt{\frac{\varphi_{-}}{z}}},\quad
M_2=-\ds{\frac{i}{2}\sqrt{\frac{\varphi_{+}}{z}}},\\[5mm]
M_3=\ds{\frac{1}{4\sqrt{2}\mu
z}\left\{\sqrt{\varphi_{+}}\left(\sqrt{\psi_{-}}\sqrt{\theta_{+}}+\sqrt{\psi_{+}}\sqrt{\theta_{-}}\right)-
\sqrt{\varphi_{-}}\left(\sqrt{\psi_{-}}\sqrt{\theta_{-}}+\sqrt{\psi_{+}}\sqrt{\theta_{+}}\right)\right\}},\\[5mm]
\alpha_1=\ds{\frac{1}{2\sqrt{2}\mu}\left(\sqrt{\psi_{+}}\sqrt{\theta_{+}}+\sqrt{\psi_{-}}\sqrt{\theta_{-}}\right)},\\[5mm]
\alpha_2=-\ds{\frac{i}{2\sqrt{2}\mu}\left(\sqrt{\psi_{+}}\sqrt{\theta_{-}}+\sqrt{\psi_{-}}\sqrt{\theta_{+}}\right)}, \\[5mm]
\alpha_3=\sqrt{z}.
\end{array}
\end{equation}

Считая, что в представленных выше формулах
выбрано $\mu \geqslant 0$, запишем условие
существования вещественных решений
\eqref{xx2_22} в виде системы неравенств
\begin{equation*}\label{xx2_23}
\begin{array}{l}
\varphi_+(z,\xi) \ls 0, \quad  \varphi_-(z,\xi)
\gs 0,
\\[3mm]
\psi_+(z,\xi) \gs 0, \quad  \psi_-(z,\xi) \ls
0,
\\[3mm]
\theta_+(z,\xi) \gs 0, \quad  \theta_-(z,\xi)
\ls 0.
\end{array}
\end{equation*}

Заметим,что подстановка $k$ через постоянную
интеграла $F$, которую по очевидным
соображениям обозначим через $f^2$, дает
\begin{equation}\label{xx2_24}
\theta_{+}\theta_{-}=[\xi-b-f+(b+f-2h)z][\xi-b+f+(b-f-2h)z]z^2.
\end{equation}
Кроме того
\begin{equation*}\label{xx2_25}
\psi_{+} \psi_{-}=\varphi_{+}\varphi_{-}=
[(z-\xi)^2+2b-k][(z+\xi)^2-2b-k].
\end{equation*}
Отсюда следует, что проекция интегрального
многообразия $J_{h,k}$ на плоскость $(z,\xi)$
ограничена сегментами (возможно, и
неограниченными) прямых
\begin{equation*}\label{xx2_26}
\begin{array}{c}
\xi-b \pm f+(b \mp f -2h)z=0, \\[3mm]
z-\xi \pm \sqrt{k-2b}=0, \qquad z+\xi \pm
\sqrt{k+2b}=0,
\end{array}
\end{equation*}
каждая из которых, если она корректно
определена, служит касательной к кривой второго
порядка
\begin{equation}\label{xx2_27}
    \mathcal{G}: \; kz^2-2b\xi z+b^2=0.
\end{equation}
Выбрав семейство таких касательных в качестве
координатной сети в соответствующей компоненте
плоскости, получим, что в такой сети для всех
допустимых пар постоянных первых интегралов
область возможности движения окажется
прямоугольной, что, как правило, ведет к
разделению переменных \cite{KhND06,khmp}.

\section{Вещественное разделение переменных}\label{sec3}
Каждая из касательных к кривой \eqref{xx2_27} в
области
\begin{equation*}\label{xx3_1}
    kz^2-2b\xi z+b^2 \gs 0,
\end{equation*}
в свою очередь, является проекцией на плоскость
$(z,\xi)$ пары прямолинейных образующих
однополостного гиперболоида \eqref{xx2_16} в
пространстве $\bR^3(z,\xi,\mu)$. Выберем в
качестве параметров двух семейств таких
образующих корни $u_1,u_2$ квадратного
уравнения
\begin{equation}\label{xx3_2}
    z u^2-2b u+(2b \xi-k z)=0.
\end{equation}
Его дискриминант, равный $\mu^2$, в силу
\eqref{xx2_11} неотрицателен на всех
траекториях, поэтому $u_1,u_2$ вещественны.
Тогда из \eqref{xx3_2}, \eqref{xx2_16} найдем
\begin{equation}\label{xx3_3}
    z=\ds{\frac{2b}{u_1+u_2}}, \qquad \xi=\ds{\frac{u_1 u_2 + k}{u_1+u_2}}, \qquad \mu=b \ds{\frac{u_1-u_2}{u_1+u_2}}.
\end{equation}

Обозначим
\begin{equation*}\label{xx3_4}
\begin{array}{l}
  p_1(u)=2b + k -u^2, \qquad p_2(u)=2b -k +u^2, \qquad p_3(u)=(u-b)^2-f^2,
\end{array}
\end{equation*}
и пусть
\begin{equation}\label{xx3_5}
    p_{ij}=p_i(u_j), \qquad r_{ij}=\sqrt{p_{ij}} \qquad (i=1,2,3;\quad j=1,2).
\end{equation}
Знак каждой из величин $r_{ij}$ произволен, но
во всех используемых одновременно формулах
должен быть выбран одинаковым. Имеем
\begin{equation*}\label{xx3_6}
\begin{array}{ll}
  \varphi_+=-\ds{\frac{p_{11}p_{12}}{(u_1+u_2)^2}} , & \varphi_-=-\ds{\frac{p_{21}p_{22}}{(u_1+u_2)^2}} , \\[5mm]
  \psi_+=\ds{\frac{ p_{12}p_{21}}{(u_1+u_2)^2}}, & \psi_-=\ds{\frac{p_{11}p_{22}}{(u_1+u_2)^2}} , \\[5mm]
  \theta_+=\ds{\frac{2b p_{31}}{(u_1+u_2)^2}} , & \theta_-=\ds{\frac{2b p_{32}}{(u_1+u_2)^2}}.
\end{array}
\end{equation*}
Подставляя эти выражения вместе с \eqref{xx3_3}
в \eqref{xx2_22}, выбираем знаки радикалов
согласованными таким образом, чтобы выполнялись
интегральные соотношения (\ref{xx2_7}). Для
этого достаточно выполнения равенства, по
смыслу согласованного с \eqref{xx2_24}:
\begin{equation*}\label{xx3_7}
    \sqrt{\varphi_+} \sqrt{\varphi_-}=\sqrt{\psi_+} \sqrt{\psi_-}.
\end{equation*}
Поэтому положим
\begin{equation*}\label{xx3_8}
\begin{array}{ll}
   \sqrt{\varphi_+}=-\ds{\ri \frac{  {r_{11}}  {r_{12}}}{u_1+u_2} } , &  \sqrt{\varphi_-}=\ds{\ri\frac{ {r_{21}}  {r_{22}}}{u_1+u_2}} , \\[5mm]
   \sqrt{\psi_+}=\ds{\frac{ {r_{12}}  {r_{21}}}{u_1+u_2}}, &  \sqrt{\psi_-}= \ds{\frac{ { r_{11}} {r_{22}}}{u_1+u_2}} , \\[5mm]
   \sqrt{\theta_+}=\ds{\frac{ {\sqrt{2b} r_{31}}}{u_1+u_2}} , &  \sqrt{\theta_-}=\ds{\frac{ {\sqrt{2b} r_{32}}}{u_1+u_2}}.
\end{array}
\end{equation*}
Тогда из \eqref{xx2_22} получаем
\begin{equation}
\label{xx3_9}
\begin{array}{l}
M_1=\ri \ds {\frac{r_{21} r_{22}}{2 \sqrt{2b
(u_1+u_2)}}},
\qquad M_2=-\ds{\frac{ r_{11} r_{12}}{2 \sqrt{2b (u_1+u_2) }}},\\[5mm]
M_3=-\ds{\frac{\ri}{2 \sqrt{b}(u_1^2-u_2^2)}}( r_{12}  r_{22}  r_{31} +  r_{11}  r_{21}  r_{32}),\\[5mm]
\alpha_1=\ds{\frac{1}{2 \sqrt{b}(u_1^2-u_2^2)}}( r_{12}  r_{21}  r_{31} +  r_{11}  r_{22}  r_{32}),\\[5mm]
\alpha_2=-\ds{\frac{\ri}{2 \sqrt{b}(u_1^2-u_2^2)}}( r_{11}  r_{22}  r_{31} +  r_{12}  r_{21}  r_{32}),\\[5mm]
\alpha_3=\ds{\frac{ \sqrt{2 b}}{
\sqrt{u_1+u_2}} }.
\end{array}
\end{equation}

Для вывода дифференциальных уравнений для новых
переменных $u_1,u_2$ из определения
\eqref{xx2_3} в силу системы \eqref{xx1_1}
найдем
\begin{equation*}\label{xx3_10}
    \dot \xi =2 \alpha_3 (\alpha_2 M_1+\alpha_1 M_2), \qquad \dot z =2 \alpha_3 (\alpha_1 M_2-\alpha_2 M_1),
\end{equation*}
что в подстановке \eqref{xx3_9} дает
\begin{equation*}\label{xx3_11}
    \begin{array}{c}
      \dot \xi = \ds{\frac{(u_1^2-k)r_{12}r_{22}r_{32}+(u_2^2-k)r_{11}r_{21}r_{31}}{\sqrt{b}(u_1^2-u_2^2)(u_1+u_2)}}, \quad
      \dot{z}=-\ds{\frac{2\sqrt{b}(r_{11}r_{21}r_{31}+r_{12}r_{22}r_{32})}{(u_1^2-u_2^2)(u_1+u_2)}}.
    \end{array}
\end{equation*}
С другой стороны, из \eqref{xx3_3} имеем
\begin{equation*}\label{xx3_12}
    \dot \xi = \ds{\frac{1}{(u_1+u_2)^2}}\left[(u_2^2-k)\dot u_1+(u_1^2-k)\dot u_2\right], \qquad \dot z = -\ds{\frac{2b }{(u_1+u_2)^2}}(\dot u_1+\dot u_2).
\end{equation*}
Из двух последних систем получим разделенную
систему
\begin{equation}\label{xx3_13}
    (u_1-u_2)\dot u_1 = \sqrt{W(u_1)}, \qquad (u_1-u_2)\dot u_2 = \sqrt{W(u_2)},
\end{equation}
где
\begin{equation}\label{xx3_14}
    W(u)=\ds{\frac{1}{b}p_1(u) p_2(u) p_3(u)}=\ds{\frac{1}{b}}(2b + k -u^2)(2b -k +u^2)[(u-b)^2-f^2].
\end{equation}
Уравнения \eqref{xx3_9}, \eqref{xx3_13} дают
полное решение задачи сведения случая
Д.Н.\,Горячева к квадратурам.

%%%%%%%%%%%%%%%%%%%%%%%%%%%%%%%%%%%%%%%%%%%%%%%%%%%%%%%%%%%%%%%%%%%%
\section{Допустимая область и бифуркационная диаграмма}\label{sec4}
Начиная с этого момента, считаем, что оставшийся свободный физический параметр положителен
\begin{equation}\label{xx4_1}
    b > 0.
\end{equation}
Полученные выше аналитические результаты по разделению переменных справедливы и для $b<0$, но качественное поведение системы может значительно отличаться. В частности, полная энергия в окрестности особенности $\alpha_3=0$ окажется неограниченной снизу.

В качестве пары независимых (почти всюду) интегралов на $\mPs$ удобно выбрать $K,F$. Ведем интегральное отображение
\begin{equation}\label{xx4_2}
    J=K\times \sqrt{F}: \mPs \to \bR^2
\end{equation}
и рассмотрим интегральные многообразия
\begin{equation*}\label{xx4_3}
    J_{k,f} = J^{-1}(k,f)=\{\zeta \in \mPs: K(\zeta)=k, \; F(\zeta)=f^2\} \qquad (f\geqslant 0).
\end{equation*}
В силу соотношения \eqref{xx2_2} они совпадают с соответствующими многообразиями \eqref{xx2_7}.

Далее мы пользуемся методами работы \cite{KhND10}. Напомним некоторую терминологию.

Допустимым множеством называется подмножество $\mD$ плоскости $(k,f)$, состоящее из всех точек, для которых $J_{k,f} \ne \varnothing$. Фиксируем $(k,f)\in \mD$. Достижимой областью $\A (k,f)$ называем проекцию интегрального многообразия $J_{k,f}$ на плоскость $(u_1,u_2)$. Из \eqref{xx3_9}, \eqref{xx3_13} в предположении \eqref{xx4_1} следует, что любая связная компонента достижимой области является прямоугольником, ограниченным отрезками прямых, параллельных координатным осям. На этих прямых значение постоянной координаты является корнем многочлена $W(u)$, называемого максимальным многочленом.

Напомним, что бифуркационной диаграммой
называется подмножество $\Sigma \subset \bR^2$,
над которым отображение \eqref{xx4_2} не
является локально-тривиальным. В условиях
рассматриваемой задачи оно совпадает с
множеством критических значений $J$. Наличие
зависимостей \eqref{xx3_9} гарантирует, что
$\Sigma$ есть часть дискриминантного множества
$\Delta$ многочлена \eqref{xx3_14},
содержащаяся в $\mD$:
\begin{equation}\label{xx4_4}
    \Sigma=\Delta \cap \mD.
\end{equation}
Из \eqref{xx3_14} получаем, что $\Delta$ состоит из трех прямых
\begin{equation*}\label{xx4_5}
    k=2b, \qquad k=-2b, \qquad f=0
\end{equation*}
и четырех парабол
\begin{equation}\label{xx4_6}
    k= \pm 2b+(f \pm b)^2.
\end{equation}

Для нахождения допустимого множества требуется получить необходимые и достаточные условия вещественности значений \eqref{xx3_9} в терминах знаков подрадикальных выражений в наиболее компактной форме независимых неравенств, легко проверяемых в областях, на которые плоскость постоянных интегралов делится дискриминантным множеством.

Пусть $\mB = \{ 0,1\} $ -- булева пара.
Следуя \cite{KhND10}, вводим функцию (булев знак) $\lsgn : \bR \to \mB$, такую, что
\begin{equation*}\label{xx4_7}
\lsgn (\theta ) = \left\{
\begin{array}{l}
{0,\quad \theta  \geqslant 0}  \\ {1,\quad \theta  < 0}
\end{array}
\right. .
\end{equation*}
Обозначая символом $\oplus$ сумму по модулю $2$, имеем свойство
\begin{equation}\notag
\lsgn (\theta _1 \theta _2 ) = \lsgn (\theta _1 ) \oplus \lsgn
(\theta _2 ).
\end{equation}

Введем булевы переменные
\begin{equation*}\label{xx4_8}
    z_i = \lsgn p_{i1}, \qquad z_{3+i} = \lsgn p_{i 2} \qquad (i=1,2,3).
\end{equation*}
При отсутствии кратных корней многочлена $W$ любая компонента достижимой области $\A(k,f)$ имеет внутреннюю точку. В такой точке условия вещественности \eqref{xx3_9} запишутся в виде системы $\mB$-линейных уравнений
\begin{equation}\label{xx4_9}
    \begin{array}{l}
      z_2 \oplus z_5=1, \\
      z_1 \oplus z_4=0, \\
      z_3 \oplus z_4 \oplus z_5 = 1, \\
      z_1 \oplus z_2 \oplus z_6 = 1, \\
      z_2 \oplus z_3 \oplus z_4 = 0, \\
      z_1 \oplus z_5 \oplus z_6 = 0, \\
      z_1 \oplus z_3 \oplus z_5 = 1, \\
      z_2 \oplus z_4 \oplus z_6 = 1.
    \end{array}
\end{equation}
Очевидно, ранг этой системы над $\mB$ равен 4 и она эквивалентна системе
\begin{equation}\label{xx4_10}
    A \mbf{z} = {\bs \zeta}^0,
\end{equation}
где $\mbf{z}=(z_1,...,z_6)^T \in \mB^6$,
булева матрица $A$ имеет вид (нуль изображаем пустой клеткой для наглядности)
\begin{equation}\label{xx4_11}
\begin{array}{c||c  c  c  c c c ||} \multicolumn{1}{c}{} &
\multicolumn{1}{c}{z_1 } & \multicolumn{1}{c}{z_2 } &
\multicolumn{1}{c}{z_3 } & \multicolumn{1}{c}{z_4 } &
\multicolumn{1}{c}{z_5 } & \multicolumn{1}{c}{z_6 } \\
{\zeta_1} & {1} & { } & { } & { }   & {1}  & {1}  \\
{\zeta_2} & { } & {1} & { } & { }   & {1}  & { }  \\
{\zeta_3} & { } & { } & {1} & { }   & { }  & {1}  \\
{\zeta_4} & { } & { } & { } & {1}   & {1}  & {1}
\end{array}\, ,
\end{equation}
а ${\bs \zeta}^0 =(0,1,1,0)^T$.

Заметим, что
\begin{equation*}
    p_1(u) > -p_2(u).
\end{equation*}
поэтому к системе \eqref{xx4_10} можно добавить условия-импликации
\begin{equation*}
\left\{\begin{array}{l}
    (z_1 \to \neg z_2)=1, \\
    (z_4 \to \neg z_5)=1,
    \end{array} \right. \Leftrightarrow \left\{\begin{array}{l}
    (z_1, z_2) \ne (1,1), \\
    (z_4, z_5) \ne (1,1).
    \end{array} \right.
\end{equation*}
Кроме того, транспозиция $u_1 \leftrightarrow u_2$ в формулах \eqref{xx3_9} равносильна замене знака у всех $r_{ij}$ одновременно, поэтому можно условиться, например, о выполнении неравенства
\begin{equation}\label{xx4_12}
    u_1^2 < u_2^2,
\end{equation}
что дает
\begin{equation*}
    p_{21}< p_{22},
\end{equation*}
откуда
\begin{equation*}
    (z_5 \to z_2)=1 \Leftrightarrow (z_2,z_5) \ne (0,1).
\end{equation*}
Перечисленным условиям отвечает лишь одно из четырех решений системы \eqref{xx4_10}
\begin{equation*}
    \mbf{z}=(0,1,1,0,0,0)^T.
\end{equation*}
Следовательно, единственная система неравенств, обеспечивающая вещественность решения \eqref{xx3_9} при договоренности \eqref{xx4_12}, имеет вид
\begin{equation}\label{xx4_13}
\begin{array}{l}
   (k+2b)-u_1^2 \gs 0, \quad u_1^2-(k-2b) \ls 0, \quad (u_1-b)^2-f^2 \ls 0, \\
   (k+2b)-u_2^2 \gs 0, \quad u_2^2-(k-2b) \gs 0, \quad (u_2-b)^2-f^2 \gs 0.
\end{array}
\end{equation}
Так как мы условились брать $f$ неотрицательным, то отсюда сразу следует, что решения возможны лишь в квадранте
\begin{equation}\label{xx4_14}
    k \gs 2b, \qquad f \gs 0.
\end{equation}
Введем в силу этого обозначения
\begin{equation}\label{xx4_15}
    \begin{array}{lll}
      m_1=\sqrt{k-2b}, & m_2=\sqrt{k+2b} & (0<m_1<m_2), \\
      n_1=b-f,  & n_2 = b+f & (n_1 \ls n_2).
    \end{array}
\end{equation}
При условии \eqref{xx4_1} из последнего уравнения \eqref{xx3_9} с необходимостью вытекает требование
\begin{equation*}\label{xx4_16}
    u_1+u_2 \gs 0.
\end{equation*}
Поэтому при выборе \eqref{xx4_12} получаем, что $u_2 \gs 0$, и тогда система \eqref{xx4_13} определяет следующие условия на достижимую область
\begin{equation}\label{xx4_17}
    \begin{array}{l}
      u_1 \in [-m_1,m_1] \cap [n_1,n_2], \qquad
      u_2 \in [m_1, m_2] \backslash [n_1,n_2].
    \end{array}
\end{equation}

\begin{figure}[ht]
\centering
\includegraphics[width=100mm,keepaspectratio]{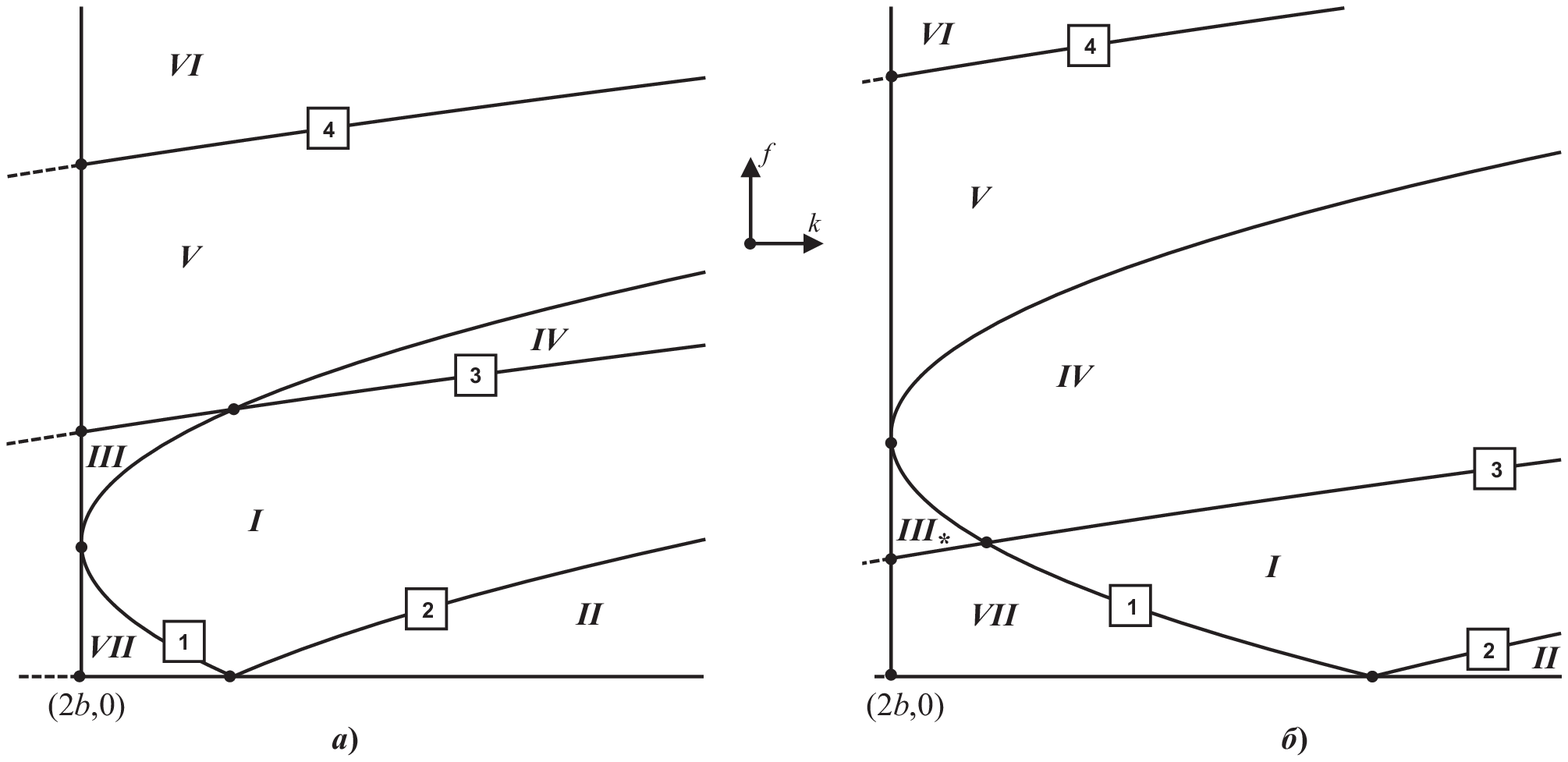}
\caption{Дискриминантное множество и кодировка областей: а) $b<1$; б) $b>1$.}\label{fig1}
\end{figure}

Пересечение множества $\Delta$ с квадрантом \eqref{xx4_14} показано на рис.~\ref{fig1}. При переходе через значение $b=1$ исчезает область $\ts{III}$ и возникает область, обозначенная $\ts{III}_*$. На параболах \eqref{xx4_6}, в соответствии с их нумерацией на рис.~\ref{fig1}, имеем:
1) $m_1=|n_1|$;
2) $m_1=n_2$;
3) $m_2=n_2$;
4) $m_2=|n_1|$.
Сводка результатов по распределению значений \eqref{xx4_15} приведена в первых двух столбцах табл.~\ref{tab1}. Оставшиеся два столбца содержат промежутки осцилляции переменных $u_1,u_2$, определенные согласно \eqref{xx4_17}.
Таким образом, движения возможны лишь в областях $\ts{I}-\ts{III}$, при этом проекция интегрального многообразия на плоскость разделенных переменных состоит из одного прямоугольника.

\begin{table}[ht]
\centering
\begin{tabular}{|c| c| c| c|}
\multicolumn{4}{r}{\fts{Таблица \myt\label{tab1}}}\\
\hline
\begin{tabular}{c}Номер\\области\end{tabular} &\begin{tabular}{c}Корни $W$\end{tabular}
&\begin{tabular}{c}Область\\изменения $u_1$\end{tabular}&\begin{tabular}{c}Область\\изменения $u_2$\end{tabular} \\
\hline
$\ts{I}$
&
$-m_2<-m_1<n_1<m_1<n_2<m_2$
&
$[n_1,m_1] $ & $[n_2,m_2] $
\\
\hline
$\ts{II}$
&
$-m_2<-m_1<n_1<n_2<m_1<m_2$
&
$[n_1,n_2] $&$[m_1,m_2]$\\
\hline
$\ts{III}$&$-m_2<n_1<-m_1<m_1<n_2<m_2$&$[-m_1,m_1] $&$[n_2,m_2] $\\
\hline
$\ts{III}_*$&$-m_2<-m_1<m_1<n_1<m_2<n_2$&$\varnothing$&$ [m_1,n_1] $\\
\hline
$\ts{IV}$&$-m_2<-m_1<n_1<m_1<m_2<n_2$&$[n_1,m_1] $&$\varnothing$\\
\hline
$\ts{V}$&$-m_2<n_1<-m_1<m_1<m_2<n_2$&$[-m_1,m_1] $&$\varnothing$\\
\hline
$\ts{VI}$&$n_1<-m_2<-m_1<m_1<m_2<n_2$&$[-m_1,m_1]$&$\varnothing$\\ \hline
$\ts{VII}$&$-m_2<-m_1<m_1<n_1<n_2<m_2$&$\varnothing$&$[m_1,n_1]\cup[n_2,m_2] $\\
\hline
\end{tabular}
\end{table}

Суммируем полученную информацию относительно допустимого множества.
\begin{theorem}
Допустимое множество в пространстве констант первых интегралов $K,\sqrt{F}$ имеет вид:

$1)$ при $b<1$
$$
\begin{array}{l}
\mD=\{0 \ls f \ls b, k \gs 2b+(f-b)^2\}\cup\{b\ls f \ls 2\sqrt{b}-b, k\gs 0\}\cup \\
\phantom{\mD=}\cup \{f \gs 2\sqrt{b}-b, k\gs -2b+(f+b)^2\};
\end{array}
$$

$2)$ при $b \gs 1$
$$
\begin{array}{l}
\mD=\{0 \ls f \ls 1, k \gs 2b+(f-b)^2\}\cup\{f \gs 1, k\gs -2b+(f+b)^2\}.
\end{array}
$$
\end{theorem}

В силу этих неравенств получим согласно \eqref{xx4_4} утверждение о бифуркационной диаграмме.
\begin{theorem}
Бифуркационная диаграмма отображения $K {\times} \sqrt{F}$ состоит из следующих множеств: при $b<1$

$a)$ $k = 2b, \quad b \ls f \ls 2\sqrt{b}-b,$

$b)$ $f=0, \quad k \gs 2b+b^2,$

$c)$ $k=2b+(f-b)^2, \quad 0 \ls f \ls 1,$

$d)$ $k=2b+(f+b)^2, \quad f \gs 0,$

$e)$ $k=-2b+(f+b)^2, \quad f \gs 2\sqrt{b}-b;$

\noindent а при $b \gs 1$

$a)$ $f=0, \quad k \gs 2b+b^2,$

$b)$ $k=2b+(f-b)^2, \quad 0 \ls f \ls 1,$

$c)$ $k=2b+(f+b)^2, \quad f \gs 0,$

$d)$ $k=-2b+(f+b)^2, \quad f \gs 1.$
\end{theorem}

Отрезок оси $f=0$ включается в бифуркационную диаграмму, поскольку очевидно, что ноль~-- критическое значение функции $F$, а для арифметического корня $\sqrt{F}$ в соответствующих точках нарушается гладкость.

\begin{figure}[ht]
\centering
\includegraphics[width=100mm,keepaspectratio]{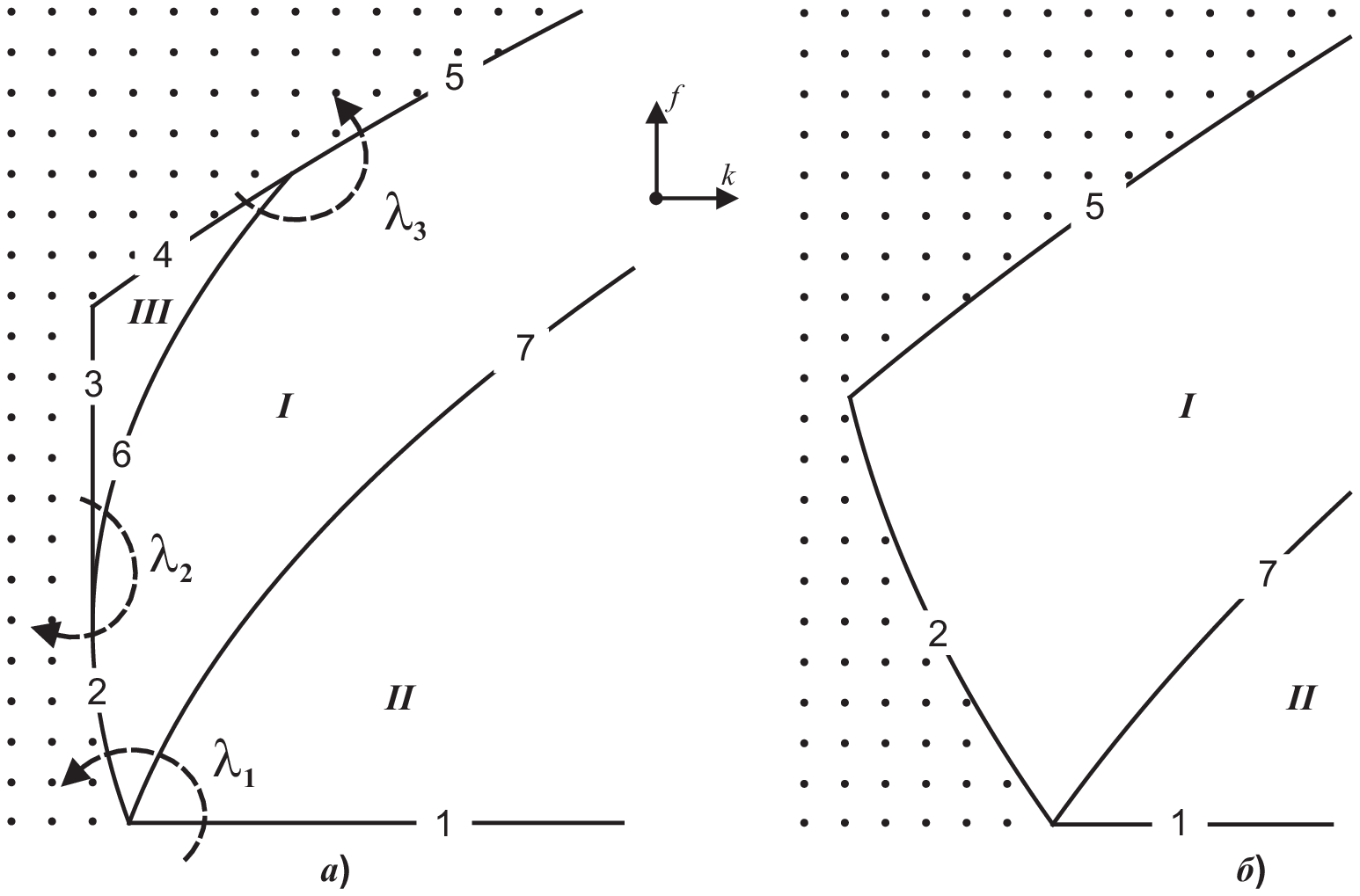}
\caption{Допустимое множество и бифуркационная диаграмма: а) $b<1$; б) $b>1$.}\label{bifset}
\end{figure}

Допустимое множество и бифуркационная диаграмма указаны на рис.~\ref{bifset}. Номерами 1--7 отмечены различные участки множеств (a)--(e), на которые их разбивают узловые точки диаграммы. Очевидно, что при переходе к значениям $b>1$ картина лишь упрощается, поэтому топологический анализ достаточно провести для случая $b<1$. На рис.~~\ref{bifset},{\it а} отмечены пути $\lambda_1 - \lambda_3$, указав бифуркации вдоль которых, мы получим полную информацию для построения любого грубого топологического инварианта.

\section{Фазовая топология}\label{sec5}
Рассмотрим наиболее богатый случай $b<1$. Для вычисления количества компонент
связности регулярных интегральных многообразий (количества торов Лиувилля) и
критических интегральных поверхностей воспользуемся методом булевых
вектор-функций \cite{KhND10}.
Сопоставим каждому радикалу $r_{ij}$ в \eqref{xx3_5} его булев знак
\begin{equation*}\label{xx5_1}
    z_i = \lsgn r_{i1}, \qquad z_{3+i} = \lsgn r_{i 2} \qquad (i=1,2,3).
\end{equation*}
Напомним, что здесь подразумеваются следующие
предварительные действия, не влияющие на
используемые формулы.

Фиксирована некоторая связная компонента $\Pi$
достижимой области  -- прямоугольник осцилляции
пары $(u_1,u_2)$. Тогда вполне определены знаки
всех подкоренных выражений $p_{ij}$ в
зависимостях \eqref{xx3_9}. У отрицательных
$p_{ij}$ изменим знак, множитель $\ri$ вынесем
в коэффициент перед произведением радикалов и
соответственно переопределим $r_{ij}$ так, что
это значение будет вещественным.

Сопоставляя каждому моному от радикалов в
\eqref{xx3_9} компоненту булевой
вектор-функции, получим те выражения для
компонент, которые фигурируют в левых частях
уравнений \eqref{xx4_9}. Получим $\mB$-линейную
функцию $\tilde A: \mB^6 \to \mB^8$. Исключение
зависимых компонент (элементарные
преобразования строк соответствующей матрицы),
сводят ее к функции $A: \mB^6 \to \mB^4$ с
матрицей \eqref{xx4_11}, которую также
обозначим через $A$. Дальнейший алгоритм
состоит в следующем. На заданном прямоугольнике
$\Pi$ аргументы функции $A$ разбиваем на две
группы, помещая в первую булевы знаки
радикалов, не меняющих знак на траекториях в
$\Pi$, а во вторую -- булевы знаки радикалов,
периодически меняющих знак на этих траекториях.
Для каждого из нефиктивных аргументов второй
группы элементарными преобразованиями матрицы
$A$ добьемся того, чтобы соответствующий
столбец стал единичным, после чего исключим из
$A$ этот столбец и строку, содержащую выбранный
аргумент. В результате получим $\mB$-линейное
отображение, зависящее только от аргументов
первой группы. Если $p$ -- его ранг, то
количество связных компонент интегрального
многообразия,  накрывающего прямоугольник $\Pi$
равно $2^p$ \cite{KhND10}. В случае, когда
компонент, зависящих только от аргументов
первой группы, не остается, полагаем $p=0$, и
соответствующий прямоугольник накрывается одной
компонентой. В критическом случае (наличие
кратного корня у максимального многочлена)
делаем то же самое с дополнительной оговоркой о
том, что радикалы, содержащие кратный корень,
всегда относятся ко второй группе, даже если
этот кратный корень -- внутренняя точка для
промежутка осцилляции и фактически за конечное
время не достигается соответствующей переменной
разделения. Приведенные рассуждения применяются
без учета возможности выбора различных знаков у
радикала $\sqrt{u_1+u_2}$. Этот выбор задает
одно из двух инвариантных подмногообразий в
$\mP$, на которые фазовое пространство системы
разбивает исключенное из рассмотрения
подмножество $\alpha_3=0$ особенностей
потенциала. Поэтому в полном фазовом
пространстве все последующие результаты о
количестве компонент связности необходимо
умножить на два. Будем считать, что для
определенности мы ограничиваемся
подмногообразием в $\mP$, заданным неравенством
$\alpha_3 >0$ (очевидно, рассматриваемая
система обладает симметрией -- обращением знака
величин $M_1,M_2, \alpha_3$).

Информация о достижимых областях, аргументах второй группы и остающихся после редукции компонентах булевой вектор-функции собрана в табл.~\ref{tab2}.

\begin{table}[ht]
\centering
\begin{tabular}{|c| c| c| c|c|c|}
\multicolumn{4}{r}{\fts{Таблица \myt\label{tab2}}}\\
\hline
\begin{tabular}{c}Номер\\области\\(сегмента)\end{tabular} &\begin{tabular}{c}Достижимая\\область\end{tabular}
&\begin{tabular}{c}Вторая\\группа\end{tabular}&\begin{tabular}{c}Редуцированная\\функция\end{tabular} & \begin{tabular}{c} $2^p$ \end{tabular} & \begin{tabular}{c} $\ds{\frac{1}{2}J_{k,f}}$ \end{tabular}\\
\hline
$\ts{I}$ & $[n_1,m_1]{\times}[n_2,m_2]$ & 1,3,5,6 & $0$ & 1 &  $\bT^2$ \\
\hline
$\ts{II}$ & $[n_1,n_2]{\times}[m_1,m_2]$ & 3,4,5 & $z_1\oplus z_2\oplus z_6$ & 2 &  $2\bT^2$ \\
\hline
$\ts{III}$ & $[-m_1,m_1]{\times}[n_2,m_2]$ & 1,5,6 & $z_2\oplus z_3$ & 2 &  $2\bT^2$ \\
\hline

$1$ & $\{n_1=n_2\}{\times}[m_1,m_2]$ & 3,4,5 & $z_1\oplus z_2\oplus z_6$ & 2 &  $2S^1$ \\
\hline
$2$ & $\{n_1=m_1\}{\times}[n_2,m_2]$ & 1,3,5,6 & $0$ & 1 &  $S^1$\\
\hline

$3$ & $\{-m_1=m_1\}{\times}[n_2,m_2]$ & 1,5,6 & $z_2\oplus z_3$ & 2 &  $2S^1$\\
\hline
$4$ & $[-m_1,m_1]{\times}\{n_2=m_2\}$ & 1,5,6 & $z_2\oplus z_3$ & 2 &  $2S^1$\\
\hline
$5$ & $[n_1,m_1]{\times}\{n_2=m_2\}$ & 1,3,5,6 & $0$ & 1 &  $S^1$\\
\hline

$6$ & $[n_1=-m_1,m_1]{\times}[n_2,m_2]$ & 1,3,5,6 & $0$ & 1 &  $S^1\vee S^1$\\
\hline

$7$ & $[n_1,m_1=n_2]{\times}[m_1=n_2,m_2]$ & 1,3,4,5,6 & $0$ & 1 & $S^1\vee S^1$\\
\hline

\end{tabular}
\end{table}

Поясним преобразования матрицы \eqref{xx4_11} в трех основных случаях.

Для области $\ts{I}$ и сегментов $2,5,6,7$ последовательно исключаем пары <<столбец-строка>> $(z_1,\zeta_1)$, $(z_3,\zeta_4)$, после чего последний столбец становится единичным и исключается пара $(z_6,\zeta_4)$. Остается одна строка, в которой аргумент $z_5$ из второй группы -- исключаем и ее. Остается вектор-функция без компонент (в табл.~\ref{tab2} она отмечена как тождественный ноль), то есть $p=0$. На сегменте $7$ тот факт, что и аргумент $z_4$ оказывается во второй группе, уже не нужен.

Для области $\ts{II}$ и сегмента $1$ последовательно исключаем пары <<столбец-строка>> $(z_3,\zeta_3)$, $(z_4,\zeta_4)$, после чего к строке $\zeta_1$ прибавим строку $\zeta_2$. Столбец $z_5$ становится единичным и исключается пара $(z_5,\zeta_2)$. Остается одна строка, отвечающая функции $z_1\oplus z_2 \oplus z_6$, в которой все аргументы из первой группы. Следовательно, $p=1$.

Для области $\ts{III}$ и сегментов $3,4$ исключаем пару $(z_1,\zeta_1)$, после чего к строке $\zeta_4$ прибавим строку $\zeta_1$. Столбец $z_5$ становится единичным и исключается пара $(z_5,\zeta_2)$. К строке $\zeta_4$ прибавим строку $\zeta_3$. Столбец $z_6$ становится единичным и исключается пара $(z_6,\zeta_3)$.  Остается одна строка, отвечающая функции $z_2\oplus z_3$ с аргументами из первой группы. Следовательно, $p=1$.

\begin{figure}[ht]
\centering
\includegraphics[width=30mm,keepaspectratio]{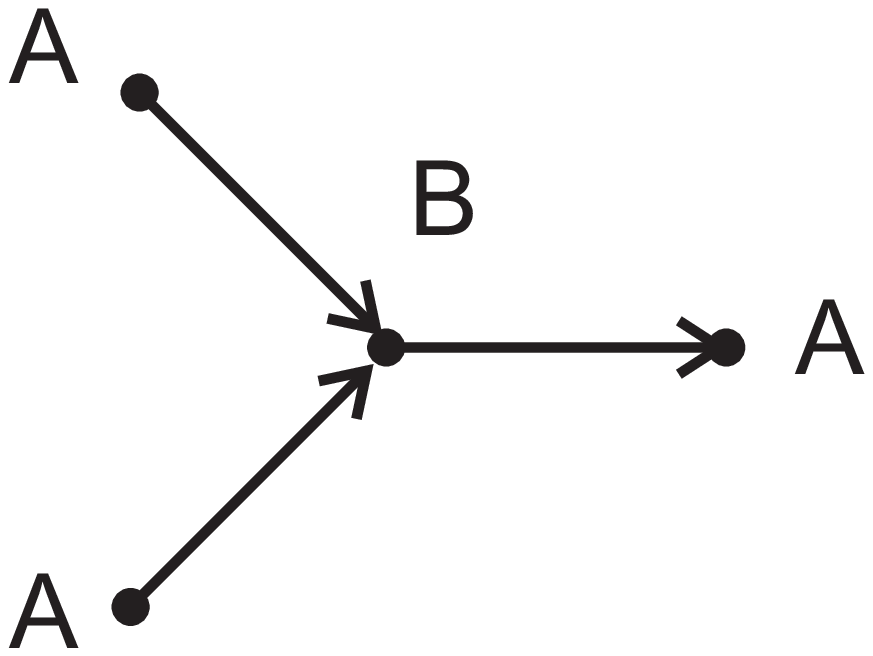}
\caption{Круговая молекула для узловых точек.}\label{molec}
\end{figure}

Теперь по количеству компонент связности однозначно определяются и интегральные многообразия, лежащие в выбранной <<половине>> системы (фиксированный знак $\alpha_3$). Они указаны в последнем столбце таблицы. Соответственно, вдоль всех путей $\lambda_1 - \lambda_3$ с учетом указанного направления имеем одинаковую последовательность бифуркаций
$\varnothing \to 2A \to 2\bT^2 \to B \to \bT^2 \to A \to \varnothing$. Молекула показана на рис.~\ref{molec}. Использованы обозначения атомов в соответствии с современной классификацией~(см., например, \cite{BolFom}).

При переходе к значениям $b>1$ происходят очевидные упрощения -- исчезает регулярная область $\ts{III}$ с двумя торами Лиувилля и примыкающие к ней сегменты $3,4,6$.

Отметим, что наличие явного алгебраического решения с разделенными переменными позволяет легко вычислить инварианты траекторной эквивалентности, основываясь на числах вращения.

Автор выражает искреннюю признательность
профессору М.П.~Харламову за полезные советы и
постоянное внимание к работе.

\end{document}